# Properties of selected mutations and genotypic landscapes under Fisher's Geometric Model

François Blanquart[1*], Guillaume Achaz[2], Thomas Bataillon[1], Olivier Tenaillon[3].

1. Bioinformatics Research Centre, University of Aarhus. 8000C Aarhus, Denmark.

2. Unité Mixte de Recherche (UMR) 7138 and Atelier de Bioinformatique, Centre National de la Recherche Scientifique (CNRS), Paris, France.

3. Institut National de la Santé et de la Recherche Médicale (INSERM), Unité Mixte de Recherche en Santé (UMR-S) 1137, IAME, F-75018 Paris, France.

*Corresponding author: francois.blanquart@normalesup.org

Correspondence:

François Blanquart, francois.blanquart@normalesup.org

Bioinformatics Research Centre, C.F. Møllers Alle 8, Building 1110, 8000C Aarhus, Denmark.

Guillaume Achaz, guillaume.achaz@upmc.fr

Thomas Bataillon, tbata@birc.au.dk

Olivier Tenaillon, olivier.tenaillon@inserm.fr

**Abstract**

The fitness landscape – the mapping between genotypes and fitness - determines properties of the process of adaptation. Several small genetic fitness landscapes have recently been built by selecting a handful of beneficial mutations and measuring fitness of all combinations of these mutations. Here we generate several testable predictions for the properties of these landscapes under Fisher's geometric model of adaptation (FGMA). When far from the fitness optimum, we analytically compute the fitness effect of beneficial mutations and their epistatic interactions. We show that epistasis may be negative or positive on average depending on the distance of the ancestral genotype to the optimum and whether mutations were independently selected or co-selected in an adaptive walk. Using simulations, we show that genetic landscapes built from FGMA are very close to an additive landscape when the ancestral strain is far from the optimum. However, when close to the optimum, a large diversity of landscape with substantial ruggedness and sign epistasis emerged. Strikingly, landscapes built from different realizations of stochastic adaptive walks in the same exact conditions were highly variable, suggesting that several realizations of small genetic landscapes are needed to gain information about the underlying architecture of the global adaptive landscape.

**Introduction**

Sewall Wright (1932) introduced the metaphor of "fitness landscapes" to think about evolutionary processes. A fitness landscape is defined by a set of genotypes, the mutational distance between them and their associated fitness. Populations are abstracted into groups of particles that navigate in this landscape (Orr 2005). In this regard, the process of adaptation by natural selection is conditioned by the fitness landscape. Many fundamental features of adaptation depend on whether the landscape is smooth or rugged, and on the level of epistasis between genotypes on the landscape (note that these two properties are related, Weinreich et al. 2005, Poelwijk et al. 2011). For examples, levels and type of epistasis determine the probability of speciation (Gavrilets 2004, Chevin et al. 2014) and the benefits of sexual reproduction (Kondrashov and Kondrashov 2001; de Visser et al. 2009; Otto 2009; Watson et al., Evolution 2011). The ruggedness of the landscape determines the repeatability and predictability of adaptation (Kaufmann 1993; Colegrave and Buckling 2005; Chevin et al. 2010; Salverda et al. 2011).

It is now possible to explore the fitness landscapes of microbial species using several experimental methods. A common type of experiment consists in isolating a number of mutants and measuring the fitness of genotypes with either a single mutation or various combinations of mutations. The most fascinating of these experiments are perhaps those considering a small number ($L$) of mutations and reconstructing all possible genotypes ($2^L$ genotypes) from the wild type to the evolved (reviewed in Weinreich et al. 2013; Lee et al. 1997, de Visser et al. 1997, Whitlock and Bourguet 2000, Lunzer et al. 2005, Weinreich et al. 2006, O'Maille et al. 2008, Lozovsky et al. 2009, da Silva et al. 2010, Chou et al. 2011, Khan et al. 2011). The properties of these reconstructed fitness landscapes determine whether adaptation was constrained to follow the particular sequence of mutations that indeed evolve in the experiment, or whether mutations could have evolved in any order with equal probability.

Various theoretical fitness landscape models have been imagined in the light of which the experimental data could be interpreted. Many models directly define the mapping between individual genotypes and fitness ("discrete" fitness landscape models). The simplest is the additive model, whereby the log-fitness is the sum of additive contributions by individual loci. This model results in no epistasis and a very smooth landscape. At the opposite extreme, the "House of Cards" model (Kingman 1978) assumes that the fitness of each genotype is drawn independently of other genotypes in some distribution. This model results in a highly epistatic and rugged landscape. In between these two extremes, two models where the roughness is a tunable parameter have been designed. The "Rough Mount Fuji" model assumes that log-fitness of a genotype is the sum of additive contributions from mutations and a House of Cards random component (Franke et al. 2011, Szendro et al. 2012). Kauffman's $NK$ model assumes that fitness results from the sum of contribution of $N$ loci, and the contribution of each locus is determined by the allelic status of this locus and $K-1$ interacting loci (often its neighbors) (Kauffman and Levin 1987, Draghi and Plotkin 2013). The contributions of these sets of loci are themselves drawn independently in some distribution. The $NK$ model encompasses all scenarios from the additive model (when $K=1$) to the full House of Cards model (when $K=N-1$). In the House of Cards, Rough Mount Fuji and NK models, epistasis is a linear combination of the random components of the fitness landscape.

A very different family of fitness landscape models specifies fitness by mapping genotypes to a set of phenotypes that are themselves under selection. The most famous of these is Fisher's geometric model (Fisher 1930). In Fisher's model, individuals are characterized by a number of continuous phenotypes that are under stabilizing selection towards a single fitness peak in the multivariate phenotypic space. Mutations fuel the process of adaptation by generating new genotypes with different phenotypic values. One fundamental difference with the models described above and phenotypic models is that in the latter, epistasis emerges from the non-linearity of the phenotype to fitness map, and not from random components.

In spite of the diversity of fitness landscape models, relatively little work has attempted to confront directly these models with experimental data. The "Rough Mount Fuji" model is able to reproduce a diversity of patterns observed in experimental fitness landscapes (Szendro et al. 2012). The $NK$ model has been little used to interpret data, perhaps because this model requires computation of a prohibitive number of fitness values when the number of loci involved in adaptation is large. In contrast, Fisher's model is sufficiently simple to allow mathematical analysis and very fast simulations, which probably explains why it has become increasingly popular to interpret experimental data. Fisher's model successfully predicts the distribution of selective coefficient of random mutations (Martin et al. 2006), levels of epistasis (Martin et al. 2007, Gros et al. 2009, Rokyta et al. 2011), levels of dominance (Manna et al. 2011), and the drift load (Tenaillon et al. 2007, Gros and Tenaillon 2009). Fisher's model has been used predominantly to interpret results on fitness effects of single mutations or pairs of mutations, and these mutations were often considered as newly arising random mutations and thus not filtered by selection. But so far no predictions have been developed for the properties of experimental genotypic landscapes under Fisher's model. Generating such predictions raises several challenges. First, the phenotypic layer between genotypes and fitness makes it is less straightforward to generate prediction for the properties of genotypic landscapes under Fisher's model than under genotypic models. Second, most experiments use mutations that arise under the action of natural selection, either naturally or in experiment (Lee et al. 1997, Sanjuan et al. 2004, Rokyta et al. 2011, O'Maille et al. 2008, Lozovsky et al. 2009, Chou et al. 2011, Khan et al. 2011). This raises a theoretical challenge because selected mutations are a non-random sample of all mutations. For example, selection has been shown to bias the prevalence and type of epistasis among mutations (Draghi and Plotkin 2013). Some experiments have used random mutations (de Visser et al. 1997, Whitlock and Bourguet 2000, Sanjuan et al. 2004), but even protocols designed to identify

"random" mutations actually involve selection at some stage (Bataillon and Bailey 2014). Third, the precise protocol used to identify mutations potentially impacts the reconstructed genotypic landscape. In some experiments, each of the mutations was selected independently in distinct populations (Sanjuan et al. 2004, Rokyta et al. 2011). In others, all mutations arose sequentially in the same population (Lee et al. 1997, O'Maille et al. 2008, Lozovsky et al. 2009, Chou et al. 2011, Khan et al. 2011). The properties of selected mutations potentially depends on the genetic background in which they arise (because of epistasis), thus of the details of the protocol.

Here we combine a new analytical approximation and simulations to address these challenges and generate predictions for the properties of genotypic landscapes under Fisher's model. We focus on selected mutations and we contrast several protocols (independently selected vs. co-selected mutations). We analyze the properties of selected mutations at several scales. We begin by introducing Fisher's fitness landscape model and derive new analytical results on the properties and fitness effect of single selected mutations. In a second step, we derive predictions for the distribution of the coefficient of epistasis and the fraction of sign epistasis among two mutations. Last, we explore the properties of genotypic landscapes that include a larger number of mutations.

## 1. The geometry of selected mutations in Fisher's model

### 1.1 The model

We use Fisher's fitness landscape model to define the relationship between genotypes, phenotypes, and fitness. We assume Gaussian stabilizing selection on a set of phenotypes. In mathematical terms, setting the optima at 0 for all traits without loss of generality, the fitness of an individual with trait vector $z$ is:

$$W[z] = \exp[-z^T S z] \quad (1)$$

where $S$ is a matrix representing the variance-covariance structure of selection. The diagonal elements determine the strength of direct stabilizing selection on each of the trait, and the off-diagonal elements represent correlative selection between traits. The evolutionary dynamics of the population in this landscape is fueled by mutations, which affect all traits to the same extent (universal pleiotropy). Specifically, the effects of each mutation on the set of traits are drawn in a multivariate normal distribution with variance covariance matrix $M$. Since both $S$ and $M$ are positive semi-definite matrices, it is always possible to find a linear transformation of the phenotypic space, ensuring that in the transformed space all traits are independent for selection ($S$ becomes a diagonal matrix) and all traits are independent and have equal variance by mutation ($M$ becomes proportional to the identity matrix) (Welch and Waxman 2005, Martin et al. 2006). Hereafter, for simplicity, we assume that in this transformed space $S$ is the identity matrix, such that selection acts with the same intensity on all traits. This assumption should not qualitatively affect our results because many important properties of the fitness landscape (distribution of selection coefficients, or of epistasis) only depends on the *sum* of the diagonal elements of the matrix describing selection in the transformed space, and not on the *variability* of these coefficients (specifically, these distributions depend on $\text{tr}(S.M) = \text{tr}(\Gamma)$ where $\Gamma$ is the diagonal matrix describing selection in the transformed space, (Martin and Lenormand 2006, Chevin et al. 2010). Consequently, the fitness function in the transformed space simplifies into $W[z] = \exp[-\sum_{i=1}^{n} z_i^2]$. The dimension of the trait vector and the matrix, $n$, represents the "complexity" of the organism in a trait space where selection acts independently along axes.

## 1.2 Effect of selected mutations on phenotypes and fitness

<u>An approximation</u>

We assume that the ancestral strain in which mutations arise is located far from the optimum relative to the size of mutations, and we develop a novel approximation to describe the properties of selected mutations arising in this ancestral strain. This approximation is suitable to understand and interpret many situations relevant to experimental evolution, where the ancestral strain is grown in a novel environment to which it is initially poorly adapted. For simplicity, we set the ancestral strain at position $\{\sqrt{-\log[W_0]}, 0, 0 \ldots, 0\}$ in the phenotypic space, where $W_0$ is the fitness of the ancestral strain. In other words we examine the idealized scenario where the ancestral strain is phenotypically perfect for $n-1$ traits but poorly matching the optimum for one trait. Note that it is always possible to rotate the original traits space so that this condition is satisfied.

We assume that the distribution of phenotypic effects of *random* mutations $\{dz_1, dz_2, \ldots, dz_n\}$ follows a multivariate normal distribution $N(0, \sigma_{mut}^2 I_n)$ ($N(.,.)$ denotes the normal distribution and $I_n$ is the identity matrix of dimension $n$). $\sigma_{mut}^2$ is the mutational variance which quantifies the effect of mutations in the phenotypic space. To understand how selection biases this distribution, we assume that the ancestral strain is sufficiently far from the optimum relative to the size of mutations that selection acts mainly along the first phenotypic axis that links the ancestral strain to the optimum ("main axis of selection"), and that all phenotypic changes along orthogonal directions cause negligible fitness changes. In mathematical terms, the selective coefficient of a mutation is linearly related to the phenotypic effect on the main axis ($s = -2z_{0,1}dz_1$, Appendix). Thus, the phenotypic effects of mutations along axes 2 to $n$, $\{dz_2, dz_3, \ldots, dz_n\}$ are distributed according to $N(0, \sigma_{mut}^2 I_{n-1})$ just as random mutations. To determine how selection impacts the distribution of phenotypic effects of mutations along the main axis of selection, we assume that the population size $N$ is large and the input of new mutations is small (i.e., $N\mu \ll 1$, where $\mu$ is the mutation rate), such that adaptation proceeds under a "strong selection – weak mutation" (SSWM) regime (Kimura 1983, Gillespie 1991). In this regime, the population is monomorphic most of the time, deleterious mutations have a negligible probability to fix in the population, and the next beneficial mutation to invade is the realization of a random drawing among the pool of beneficial mutations where each beneficial mutation has a probability proportional to its selection coefficient to be chosen (Patwa and Wahl 2008). Under this regime we show (Appendix) that the scaled phenotypic effect along the first trait $-d\, z_1/\sigma_{mut}^2$ is distributed according to a $\chi_2$ distribution (here and in the following $\chi_i$ denotes the chi distribution with $i$ degrees of freedom).

The geometry of selected mutations

It is more intuitive to translate these algebraic results in geometrical terms. A mutation and its fitness effect is characterized by its norms $\|z\|$ and its angle with the main axis of selection $\theta$ (see fig 1). We found that selected mutations are slightly larger than random mutations. In mathematical terms, the norm of selected mutations (the size of the mutation in the phenotypic space) scaled by the mutational standard deviation $\sigma_{mut}$ is distributed according to a $\chi_{n+1}$ while that of random mutations is a $\chi_n$. To compare properties of selected mutations and genotypic landscapes across several complexities of the phenotypic space, we scale the mutational variance to keep the norm of *random* mutations constant across complexities. More precisely we scaled the mutational variance as $\sigma_{mut} = \overline{\|z\|}_{rand} * \Gamma\left(\frac{n}{2}\right)/\left(\sqrt{2}\Gamma\left(\frac{n+1}{2}\right)\right)$, where $\Gamma(.)$ is the gamma function and $\overline{\|z\|}_{rand}$ is the average norm of a random mutation (constant across complexities). This relationship converges for large complexity to $\sigma_{mut} = \overline{\|z\|}_{rand}/\sqrt{n - \frac{1}{\sqrt{2}}}$. Note that to keep an average fitness effect of mutations arising at the optimum constant across complexities, one needs a very similar scaling $\sigma_{mut} = \bar{s}_{opt}/\sqrt{n}$. As a consequence of this scaling, the major effect of increasing complexity is to reduce the *variance* of the norm (fig. 1). We also found that selected mutations point in the direction of the optimum when complexity is low, but that they point increasingly in a direction orthogonal to the optimum as complexity increases (the distribution of $\theta$ becomes concentrated around $\pi/2$), such that mutations pointing directly to the optimum ($\theta = 0$) become extremely rare (Appendix, fig. 1). This change in orientation at high complexity is due to the overwhelming importance of other phenotypic directions relative to the "main axis of selection". In complex organisms, because of pleiotropy, beneficial mutations cause small changes on a myriad of other phenotypes as a side effect of changing the phenotypic value on the "main axis of selection". This effect of complexity was found before for random mutations, although in this context the distribution of $\theta$ is symmetrical relative to $\pi/2$ because both beneficial and deleterious mutations are present (Poon and Otto 2000).

Distribution of fitness effects of selected mutations

The properties of selected mutations imply that the (normalized) fitness effects of selected mutations are distributed as a $\chi_2$ (see Appendix). The most frequent mutations to evolve have an intermediate fitness effect, because they represent the best compromise between occurring frequently and enjoying a high selection coefficient (Kimura 1983). More precisely, under the SSWM regime and assuming the wild type is far from the optimum, we obtained the following distribution for the selection coefficient of selected mutations:

$$f^*(s) = \frac{s}{4\sigma_{mut}^2(-\log[W_0])} e^{\frac{-s^2}{8\sigma_{mut}^2(-\log[W_0])}} \text{ if } s > 0$$

$$f^*(s) = 0 \text{ otherwise}$$

(2)

The density is essentially the distribution of fitness effects of *random* mutations, weighted by their probability of fixation (assumed to be proportional to the selective coefficient $s$ in the SSWM regime). The mean of this distribution is $\sqrt{2\pi}\sigma_{mut}\sqrt{-\log[W_0]}$ and the standard deviation is $\sqrt{2(4-\pi)}\sigma_{mut}\sqrt{-\log[W_0]}$. Under our assumptions, the distribution of fitness effects does not directly depend on complexity (although it would depend indirectly on complexity through the scaling $\sigma_{mut} = \overline{\|z\|}_{rand} * \Gamma\left(\frac{n}{2}\right)/\left(\sqrt{2}\,\Gamma\left(\frac{n+1}{2}\right)\right)$ ). Indeed only the main axis of selection determines fitness, so the number of other phenotypic directions (assumed to be neutral) does not matter. Interestingly the coefficient of variation of the selection coefficient is $\sqrt{(4-\pi)/\pi} \approx 0.52$ and is independent of initial fitness and the effect of mutations in the phenotypic space.

This analytical result was found to be a good approximation when compared with the results of stochastic, individual-based simulations, when the ancestral strain is not too close to the optimum (fig. 2). These simulations model the dynamics of a population under selection, mutation (at rate $\mu = 10^{-9}$) and genetic drift (population size $N = 10^7$) under Fisher's model.

Another approximation for the distribution of fitness effects of selected mutations in Fisher's model based on a beta distribution has been proposed (Martin and Lenormand 2008). In contrast to ours, this approximation works best when the ancestral strain is very close to the optimum, such that beneficial mutations are very rare (the approximation is based on extreme value theory). The beta approximation implies, just as our $\chi$ approximation, that the most abundant selected mutations are those of intermediate fitness effect. In contrast to our approximation, the beta distribution does not depend on the phenotypic effects of mutation $\sigma_{mut}$ but depends strongly on the complexity of the organism $n$ (Appendix, equation S3d). This approximation was found to perform poorly for $W_{00} = 0.9$ (it becomes accurate when the ancestral strain is much closer to the optimum, $W_{00} = 0.99$ with $\overline{\|z\|}_{rand} = 0.1$). Lastly, we compared these approximations to the gamma approximation proposed by Martin and Lenormand (2006) adapted to the case of selected mutations, and found the gamma approximation performs best (fig. 2; Appendix, equation 3c).

Given our analytical results on the properties of individual selected mutations in Fisher's model, our objective is now to determine the emerging properties of genotypic landscapes composed of several mutations. Towards this goal, we next investigate the properties of pairs of mutations.

## 2. Properties of pairs of mutations in Fisher's landscape

2.1 Epistasis among selected mutations

We first examine the distribution of the epistasis coefficient, which quantifies non-multiplicative interactions for fitness between two mutations. The epistasis coefficient between two mutations is defined as $e = \log\left[\frac{W_{11}W_{00}}{W_{01}W_{10}}\right]$ where $W_{11}$ is the fitness of the double mutant and $W_{01}, W_{10}$ are that of the two single mutants (the fitness of the ancestral strain is now denoted $W_{00}$ for clarity, because it bears the "0" allele at two loci). In Fisher's fitness landscape model, it can be shown that epistasis is proportional to the scalar product of the effects of the two mutations in the phenotypic space (Martin et al. 2007):

$$e = -2 \sum_{i=1}^{n} dz_i dz_i' \qquad (3)$$

where $dz_i$ and $dz_i'$ are the phenotypic effect of two mutations on trait $i$. It has been shown previously that the coefficient of epistasis of random (newly arising) mutation does not depend on the fitness of the ancestral strain and is distributed as a $N(0, 4n\sigma_{mut}^4)$ (Martin *et al.* 2007). Here we derive similar results for the distribution of the epistasis coefficient between *selected* mutations.

Epistasis between independently selected mutations

First we investigate the distribution of epistasis among mutations that evolved in independent replicates. Specifically, we assume that the experiment starts with a monomorphic population where fitness is $W_{00}$ (hereafter the ancestral strain), that it is composed of a series of independent replicates evolving in parallel, and that each replicate is let to evolve until one mutation arises and fixes.

The epistasis coefficient (equation 3) can be decomposed as the sum of a "selected epistasis" component which emerges from selection along the main axis of selection (first axis), and an independent "random epistasis" component, contributed by all other orthogonal axes and which we characterize using the approximation developed by Martin *et al.* (2007) (details in Appendix). We find, under the SSWM regime and assuming the wild type is far from the optimum, that the mean and variance of the distribution of epistasis are:

$$\text{E}[e] = -\pi \sigma_{mut}^2$$

$$\text{V}[e] = (4n + 12 - \pi^2)\sigma_{mut}^4$$

(4)

Note that $12 - \pi^2 \approx 2$ so $\text{V}[e]$ quickly approaches $4n\sigma_{mut}^4$ as complexity increases. These results reveal that the average epistasis is negative, meaning that two independent mutations that both bring the population closer to the optimum along the first axis tend to interact negatively for fitness. The variance of the distribution of epistasis among selected mutations is very similar to the variance among random mutations ($\text{V}[e]_{random} = 4n\sigma_{mut}^4$ (Martin *et al.* 2007)). As $\sigma_{mut}$ is proportional to $1/\sqrt{n}$ when $n$ is large, both the average and variance of epistasis become proportional to $1/n$. The difference in the distribution of epistasis between selected and random mutations becomes marginal as complexity increases because effectively neutral phenotypes contribute increasingly more to epistasis. Importantly, epistasis among selected mutations does not depend on the fitness of the ancestral strain, just as for random mutations.

In general, our analytical approximations capture correctly the average and variance of epistasis when the population is initially not too close of the optimum (fig. 3, 4). However, slight discrepancies with the predictions arise when the ancestral strain is very close to the optimum. Epistasis among independently selected mutations tends to be more negative than predicted. Near the optimum, the selected mutations are more constrained to follow the direction of the optimum ($\theta$ is closer to 0 than expected) and this causes more negative epistasis on average. This "canalization" also causes a reduction in variance of epistasis close to the optimum (fig. 4).

Epistasis between co-selected mutations

Next we determine the distribution of epistasis arising between two mutations that arise and sweep to fixation sequentially in the same ancestral strain ("co-selected mutations"), using the individual based simulations described above. We find that co-selected mutations have an epistasis coefficient very similar to that of independently selected mutations, except when the ancestral strain is very close to the optimum, in which case the epistasis coefficient between co-selected mutations becomes much more positive than in independently selected mutations (fig. 3, fig. 4). Co-selected mutations have epistasis distribution very similar to independently selected mutations for two reasons: first, because the ancestral strain is sufficiently far from the optimum, the approximation that only the first phenotypic axis is important for selection still holds. Second, as epistasis is independent of the ancestral strain's fitness, the fact that the second mutation emerges in a background with higher fitness does not affect the distribution. However, as the ancestral strain becomes closer the optimum, the distribution of epistasis shifts towards more positive values on average than those predicted for independently selected mutations (fig. 3, 4). Indeed, close to the optimum, changes in the phenotypic directions orthogonal to selection are not neutral (as assumed in the approximation) but selected against. A selected mutation will typically bring the population closer to the optimum along the main direction of selection at the cost of antagonistic pleiotropy in the other traits. Antagonistic pleiotropy will be larger when the complexity is higher. When a second mutation arises in the background of the first, it will typically compensate for these antagonistic effects. Accordingly, the mutations typically present positive epistasis. When complexity $n$ is high, many traits are available for that compensation effect to operate and positive epistasis becomes more pervasive (fig. 3, 4). The observation that epistasis between selected mutations is negative for low complexity but positive for large complexity has been noted previously (Chevin et al. 2014).

The results developed in this part represent a step forward for evaluating the power of Fisher's fitness landscape model to explain data on the fitness effects and epistasis of selected mutations. The epistasis coefficient is important to predict the dynamics of adaptation, but is more difficult to relate to properties such as the roughness and accessibility of the underlying genotypic landscape. Such properties are determined by sign epistasis – the fact that mutations are beneficial in some background but deleterious in other backgrounds (Weinreich *et al.* 2005), which is not directly related to the epistasis coefficient. In the following part, we investigate the fraction of sign epistasis among selected mutations in Fisher's model.

## 2.2. Sign epistasis among selected mutations

For independently selected mutations, we find that a mutation will present sign epistasis in the ancestral background vs. the background with another mutation if and only if (Appendix):

$$s + e < 0 \tag{5a}$$

where $s$ is the selection coefficient of the sign epistatic mutation in the ancestral background and $e$ is the epistatic coefficient between the sign epistatic mutation and the other mutation. In the case of selected mutations, the selection coefficient $s$ is always positive, thus sign epistasis occurs when the epistasis coefficient is sufficiently negative. Clearly, this condition will be increasingly easier to fulfill as the ancestral strain is closer to the optimum because the selection coefficient becomes smaller on average, and epistasis become more shifted towards negative values.

For co-selected mutations, we find that the first mutation is sign epistatic if and only if it fulfills condition (5a), and the second mutation is sign epistatic if and only if:

$$-e + s_{(2)} < 0 \qquad (5b)$$

where $s_{(2)}$ is the selection coefficient of the second mutation (that is, $s_{(2)} = \log(W_{11}) - \log(W_{10})$). Because the epistasis coefficient is more positive on average among co-selected mutations, the first mutation should more rarely be sign epistatic among co-selected than among independently selected mutations, and thus sign epistasis should be in general less frequent among co-selected mutations. Lastly, the effect of complexity on the fraction of sign epistasis is harder to predict, because both the selection coefficient and epistasis tend to be concentrated around 0 as complexity increases.

Our predictions are verified in the simulations: sign epistasis is much more frequent when the fitness of the ancestral strain is higher, both in independently selected and co-selected mutations (fig. 5). Moreover, we found that sign epistasis is more frequent in a more complex phenotypic space. As much as 20% of sign epistasis, (12% simple and 8% reciprocal sign epistasis), occurred between independently selected mutations when $W_{00} = 0.95$ and $n = 100$. This large fraction of sign epistasis may seem surprising since Fisher's model is a completely smooth phenotypic landscape. Even more strikingly, this large fraction of sign epistasis emerges in the absence of optimum overshooting. Optimum overshooting - the fact that individual mutations are so large that their combined effects are deleterious (fig. 6) - can generate sign epistasis in Fisher's model, but is not present for the values of parameters we chose because the ancestral strain is always too far away from the optimum relative to the size of mutations. Specifically, the distance to the optimum expressed in units of average norm of mutations, which gives a rough estimate of the number of mutations that may fix in the ancestral strain until the optimum is reached, ranges from $\sqrt{-\log[W_{00}]/0.1} \approx 2$ when $W_{00} = 0.95$, up to 15 when $W_{00} = 0.1$. To understand what causes sign epistasis, we looked more specifically at the properties of sign epistatic mutations and found that these mutations had larger norm, smaller selection coefficient and tend to be more orthogonal to the main direction of selection ($\theta$ is closer to $\pi/2$). In other words, sign epistatic mutations are mutations of very small fitness effect with large antagonistic pleiotropy. Such mutations have a small beneficial effect in the background in which they evolve, but they easily become deleterious in another background (fig. 6). In more complex organisms, there are much more phenotypic axes in which antagonistic pleiotropy can act, explaining why sign epistasis is more frequent (in geometrical terms $\theta$ is more frequently close to $\pi/2$ in more complex organisms, fig. 1). Finally, note that we scale the mutational effect such that complexity does not affect the expected norm of mutations in the phenotypic space. If the scaling between mutational effect and complexity is such that the norm of mutations is larger in a more complex space, an even greater impact of complexity on sign epistasis is expected.

To conclude, a smooth phenotypic landscape such as that of Fisher's model may generate high amounts of sign epistasis among mutations because of antagonistic pleiotropy, and especially so in complex organisms. Thus, there is the potential for Fisher's landscape to generate quite rough genotypic landscapes. In the following we explore this possibility by considering genotypic landscapes composed of a larger number of mutations.

## 3. Emergence of a diversity of genotypic fitness landscape under Fisher's phenotypic model

Several experiments approach the properties of the fitness landscape by examining the genotypic landscape between an ancestral genotype to an evolved genotype differing by a small number of mutations $L$. Again these mutations may be independently selected or co-selected. The genotypic landscape is made up of $2^L$ genotypes with all possible combinations of these mutations. To quantify more precisely the roughness of these genotypic landscapes, we use simulations to examine the distribution of two statistics that have been previously used to characterize genotypic landscapes, as a function of $W_0$ and the complexity $n$, when mutations are independently selected or co-selected.

### 3.1 Statistics summarizing the properties of the genotypic landscape

We use two commonly used statistics to summarize the properties of the genotypic landscape. The first is the fraction of sign epistasis among all pairs of genotypes separated by two mutations (as already investigated for the simpler case of two mutations in part 2.2 above). This proportion is 0 in an additive landscape, and simulations show it reaches 70 to 90% on average among independently selected mutations in a House of Cards model (a very rugged genotypic landscape). The second statistic is the roughness to slope ratio (Carneiro and Hartl 2010, Szendro et al. 2012). This measure quantifies how well the landscape can be described by a linear model where mutations additively determine fitness. Specifically, the linear model is:

$$W^{(fit)} = c_0 + \sum_{j=1}^{L} c_j a_j \quad (6)$$

where the sum is over all $L$ loci, $c_j$ is the effect of the mutation at the $j$th locus on fitness, and $a_j$ is an indicator variable which is 0 or 1 if the $j$th locus is wild type or mutated respectively. The $c_j$ are estimated by least square regression. The slope is the defined as the average additive effect of mutations, $s = \frac{1}{L}\sum_{j=1}^{L}|c_j|$, and the roughness is the residual variance that quantifies the fit of the linear model, $r = \sqrt{2^{-L}\sum_{genos}(W - W^{(fit)})^2}$. The roughness to slope ratio $r/s$ is 0 when the fit of an additive model is perfect, and becomes very large in a House of Cards model.

## 3.2 Distribution of the statistics

In a first step we investigated the distribution of the two statistics when starting from an ancestral strain with fitness $W_{00} = 0.1$, and when selecting five mutations according to the two sampling protocols described above (either mutations occurring independently in different replicates, or co-selected). We found that Fisher's fitness landscape generates a genotypic landscape very close to a smooth, additive landscape when the ancestral strain is far from the optimum relative to the size of mutations ($W_{00} = 0.1$ and $\overline{\|z\|}_{rand} = 0.1$). Specifically, there is no sign epistasis in more than 95% of landscapes and the roughness to slope ratio is always very close to $0$. Interestingly, the roughness-to-slope ratio informs on the complexity of the fitness landscape (fig. 7): the genotypic landscape is closer to additive in phenotypic landscapes of higher dimension (lower $r/s$ ratio). This is because the deviation from strict additivity is due to the curvature of the fitness landscape, which appears smaller in more complex landscapes because the effects of mutations in the direction of the optimum are smaller.

In a second step, we investigated fitness landscapes that emerge when the ancestral strain is fitter ($W_{00} = 0.9$). In this case the distance to the optimum in units of average norm of mutations is approximately equal to 3. Three general tendencies emerge from the distribution of these statistics (fig. 7). First, complexity does not affect much the distribution of the statistics. Again, this is because we scale the norm of mutations such that the expected norm is constant across complexity. Second, the properties of the genotypic landscapes depend on the sampling protocol. Specifically landscapes generated from independently selected mutations tend to be more rough than those generated from co-selected mutations. Indeed, landscapes with co-selected mutations include at least one evolutionarily accessible path, and tend to exhibit mutations of smaller effect on average, which generates less rough genotypic landscapes. Third, and more strikingly, the stochasticity of the adaptive process generates a great diversity of landscapes with the same parameter values (fig. 7, 8). Fig. 8 shows several genotypic landscapes with a diversity of levels of roughness and accessibility that are generated by independent replicate simulations with the same set of parameters. In some cases, the genotypic landscape generated by independently selected mutations reflects quite clearly the presence of an optimum in fitness.

# Discussion

Main results:

Fisher's fitness landscape is based on a smooth continuous phenotypic landscape. Yet non-trivial properties of individual random mutations and their interactions emerge from that model. In the present paper, we have explored the properties of genotypic fitness landscapes generated by adaptive mutations in Fisher's fitness landscape model and developed analytical solutions when the ancestral strain is far from the optimum.

First, selected mutations may point in the direction of the optimum in simple organisms, but in more complex organisms they are often almost orthogonal to the main direction of selection because of the numerous pleiotropic effects of beneficial mutations. The fitness effects of selected mutations follow a $\chi$ distribution independent of the complexity, with an invariant coefficient of variation of approximately $0.5$.

Second, the epistasis coefficient among pairs of selected mutations is on average negative when the ancestral strain is far from the optimum. When the ancestral strain is close to the optimum, the epistasis coefficient is on average more strongly negative between independently selected mutations. For co-selected mutations, epistasis is negative on average in simple organisms but can become positive when the organism is very complex. Sign epistasis – the fact that a mutation may be beneficial or deleterious depending on the background in which it appears– may be common in Fisher's model, especially when the ancestral strain is close to the optimum relative to the size of mutations. The cause of sign epistasis in Fisher's model under our conditions is antagonistic pleiotropy, whereby the combination of pleiotropic effects in multiple phenotypic directions are deleterious in the double mutant. Antagonistic pleiotropy happens more frequently in more complex organisms.

Third, we explore how these properties of pairs of mutations scale up to the global properties of genotypic landscapes made of all combinations of mutations between an ancestral strain and an evolved genotype. When the ancestral strain is far from the optimum, these empirical landscapes are smooth and very similar to an additive landscape. However, when the ancestral strain is close to the optimum, the landscape can encompass some roughness, especially when mutations have been independently selected. Even though all landscapes have a major additive component, an appreciable variety of empirical landscapes can be observed across different replicate simulations using the same set of parameters.

It is worth noting, first that it is not so much the value of the initial fitness that matters for our qualitative results, but rather the distance to the optimum in number of mutation steps. Second, that the quantitative results presented rely on the assumption of universal pleiotropy. Most qualitative results will hold if pleiotropy were not universal, as they are based on the geometry of Fisher's model, but predictions on the quantitative impact of restricted pleiotropy or any other alteration of pleitropy may be more difficult.

Comparison with previous theoretical work:

This work complements previous work. Draghi and Plotkin (2013) studied the patterns of epistasis along adaptive walks in Kauffman's NK model. They found a predominance of antagonistic (negative) epistasis in the early steps of adaptation and of synergistic (positive) epistasis later on. This result is very similar to our finding that co-selected mutations tend to interact negatively for fitness far from the optimum, but positively close to the optimum (fig. 3). Both illustrate the limited number of options left for adaptation when close the optimum. Draghi and Plotkin's observation results directly from the finiteness of the genotypic space. Close to the optimum, there are very few beneficial mutations; if one of these mutations fixes, it is very likely that adaptation will proceed further through other beneficial mutations that were "unlocked" by the fixation of the last beneficial mutation. In Fisher's model, in contrast, more positive epistasis close to the optimum is due to the fact that most beneficial mutation entail a cost in alternative directions. The second mutation compensates for antagonistic pleiotropy of the first mutation. In our model this effect is observed in particular in more complex organisms where compensation operates in multiple phenotypic directions.

Testing our predictions with data:

Our analytical and simulation results generate a number of testable predictions for experimental work. We tested several of these predictions using published data and find that in general this data is in good agreement with Fisher's model.

First, we tested the prediction that the coefficient of variation of the distribution of fitness effect of selected mutation is approximately $0.5$ (equation 2). Two data sets on adaptation of *E. coli* to minimum glucose medium (Rozen et al. 2002) and adaptation of *Pseudomonas aeruginosa* to rifampicin (MacLean et al. 2010) are suitable to test this prediction. We found the distribution of fitness effects of selected mutations in these datasets has $c_v = 0.44$ and $c_v = 0.42$ respectively, which is in relatively good agreement with the prediction.

Second, our results on the distribution of epistasis among pairs of selected mutations suggest that complexity of the organism and the mutational variance could be estimated from data. This approach has been suggested using the distribution of fitness effects of random mutations (Martin *et al.* 2007). Yet with random mutations, $n$ and $\sigma_{mut}$ could not be estimated independently. Here on the contrary, the mean and variance of the distribution give access to $n$ and $\sigma_{mut}$ independently. For example, the coefficient of variation of epistasis is equal to $-\sqrt{\frac{4}{\pi^2}n + \frac{12}{\pi^2} - 1}$, from which the complexity $n$ can be inferred. Moreover, the distribution of the selection coefficient can be used to infer the initial fitness $W_{00}$ (equation 2). We computed $n$ and $\sigma_{mut}$ for Methylobacterium adapting to methanol medium (Chou et al. 2011). In that dataset, the strain was initially extremely maladapted and four beneficial mutations were quickly selected. Using the distribution of epistasis among these pairs of mutations we found $n = 8$ and $\sigma^2_{mut} = 0.007$. These estimates should, however, be taken with caution as our analytical results (as well as those of Martin et al. 2007) rely on a number of assumptions which may not always be fulfilled (in our case, universal pleiotropy, SSWM regime, ancestral strain far from the optimum).

Third, we tested whether the distribution of statistics characterizing the full genotypic landscape was compatible with the predictions of Fisher's model. We developed three global predictions: first mutations arising in an ancestral strain far from the optimum relative to the size of mutations generate a close-to-additive fitness landscape, while mutations arising in a fitter background generate more rough landscapes. To test this prediction, we compared the fraction of sign epistasis and roughness to slope ratio for two published experimental landscapes (Chou et al. 2011, Khan et al. 2011) that differ in their initial maladaptation as the first one exhibits an increase in fitness of 100% over 600 generations while the other one fitness increased of 30% over 2000 generations. As expected the landscape starting from a better adapted strain (Khan et al. 2011) exhibited more sign epistasis and a higher roughness to slope ratio than the landscape starting from a very poorly adapted strain (Chou et al. 2011) (fig. 7). The second prediction is that co-selected mutations tend to show less epistasis than independently selected mutations. Chou et al.'s dataset is also comforting this prediction. While mutations arising in the same replicate exhibited little sign epistasis (Chou et al. 2011), mutations occurring in independent populations exhibited substantial levels of antagonistic and sign epistasis (Chou et al. 2014). In this system all mutations resulted in a decrease of the expression of an operon carried on a plasmid. The combined effects of two mutations occurring independently resulted in too low levels of expression, hence lower fitness. In addition, the genotypic landscape reconstructed in Weinreich et al. (2006), which comprises 5 mutations which did not all evolve together (only some combinations of these mutations were found together in the same population) exhibits a high fraction of sign epistasis and roughness to slope ratio compatible with independently selected mutations in Fisher's model. It is encouraging to note that the joint values of the two statistics in the three experimental landscapes fall squarely in the density of points generated with Fisher's model on fig. 7. However the relationship between the two statistics could be a general property of fitness landscapes and not a specific prediction of Fisher's model. Indeed Szendro et al. (2012) find a similar relationship in the "Rough Mount Fuji" model (see their fig. 5). The third prediction on genotypic landscapes under Fisher's model is that when close to the optimum or when mutations have effects large enough to get close to the optimum, a very large diversity of landscapes may emerge. We do not have now enough landscapes built in similar conditions to test that prediction, but we should keep in mind that stochasticity may be a major actor in shaping those landscapes and that deterministic interpretation of their fundamental meaning and differences should be taken with care.

An important implication of our work is that small genetic landscapes built from adaptive walks are not very informative about the underlying structure of the underlying landscape. The presence of sign epistasis or ruggedness can be very contingent on the sample of mutations used to build the landscape. Over the last years, experimental evolution has shown that hundreds of beneficial mutations may appear within a single gene (Salverda et al. 2010, Tenaillon et al. 2012). This vast number of mutations comforts the relevance of continuous (phenotypic) landscapes and suggests that any landscape built form a handful of mutations is merely a single stochastic realization among of the range of possible combination of adaptive mutations. A landscape as simple as FGM yields a diversity of small genetic landscapes depending on the regime used to sample mutations, the distance to the optimum and the dimensionality of the phenotypic space. In particular when close to the optimum, under the same exact set of parameters, very different empirical landscapes may emerge as independent realizations of the stochastic sampling of a single landscape.

To summarize, the good agreement between our qualitative predictions and experimental data, and the diversity of landscapes that Fisher's model may generate, suggest that Fisher's model may be used as a flexible tool to describe the relationship between genotypes and fitness. Specifically, although Fisher's model appears very smooth at the phenotypic level, it may actually be used to interpret experimental evolution results where very short adaptive walks and highly rugged genotypic landscapes are observed (e.g., Gifford *et al.* 2011). These limitations call for more flexible and more robust methods to estimate the parameters of Fisher's model from experimental data. Of course, it would also be necessary to compare more rigorously the explanatory power of different fitness landscape models. For example, the "Rough Mount Fuji" model (Szendro et al. 2012) or the NK model (Franke et al. 2010) are also able to explain a diversity of patterns observed in experimental data. The analytical and simulation results presented here are a step forward towards this goal. But Fisher's model appears as a good candidate to standardize and unify a growing and disparate body of experimental work.

**Acknowledgements:**

F.B. and T.B acknowledge support from the Danish Research Council (FFF-FNU) and by the European Research Council under the European Union's Seventh Framework Program (FP7/2007-2013)/ERC Grant 311341 to T.B.. F.B. also benefited from a PhD grant of French "Ministère de la Recherche" and support from Bettencourt Foundation. GA is supported by the "Agence Nationale de la Recherche" through the grant ANR- 12-JSV7-0007.  O. T. was supported by the European Research Council under the European Union's Seventh Framework Program (FP7/2007-2013)/ERC Grant 310944.

# Figures

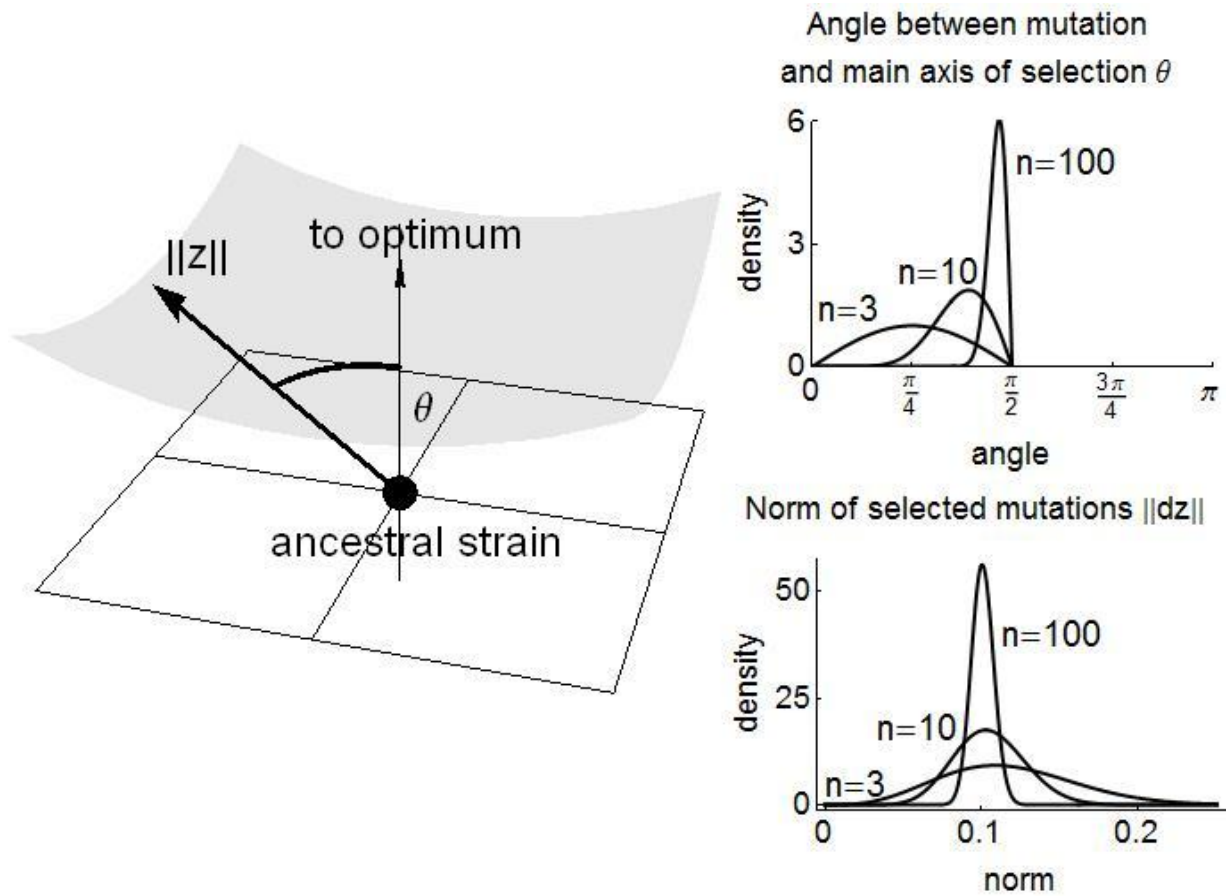

Figure 1. The geometry of selected mutations in Fisher's fitness landscape for various complexities of the phenotypic space. Left panel: a beneficial mutations is represented in a 3-dimensional space, where the ancestral strain is shown as a black point, the mutation evolving in the ancestor as plain arrows, a fitness isocline as a gray sphere, and the vertical axis is the main axis of selection. The geometry of the mutation is characterized by the norm $\|z\|$ and by the angle between the mutations and the main axis of selection $\theta$. On the right panel, the distribution of these quantities is shown for complexities $n = 3$, $n = 10$, $n = 100$. The mutational variance $\sigma_{mut}$ was normalized such that the expected norm is the same for all complexities. At higher complexities, mutations tend to be almost orthogonal to the main axis of selection ($\theta \to \pi/2$) and to exhibit very little variation in their norm.

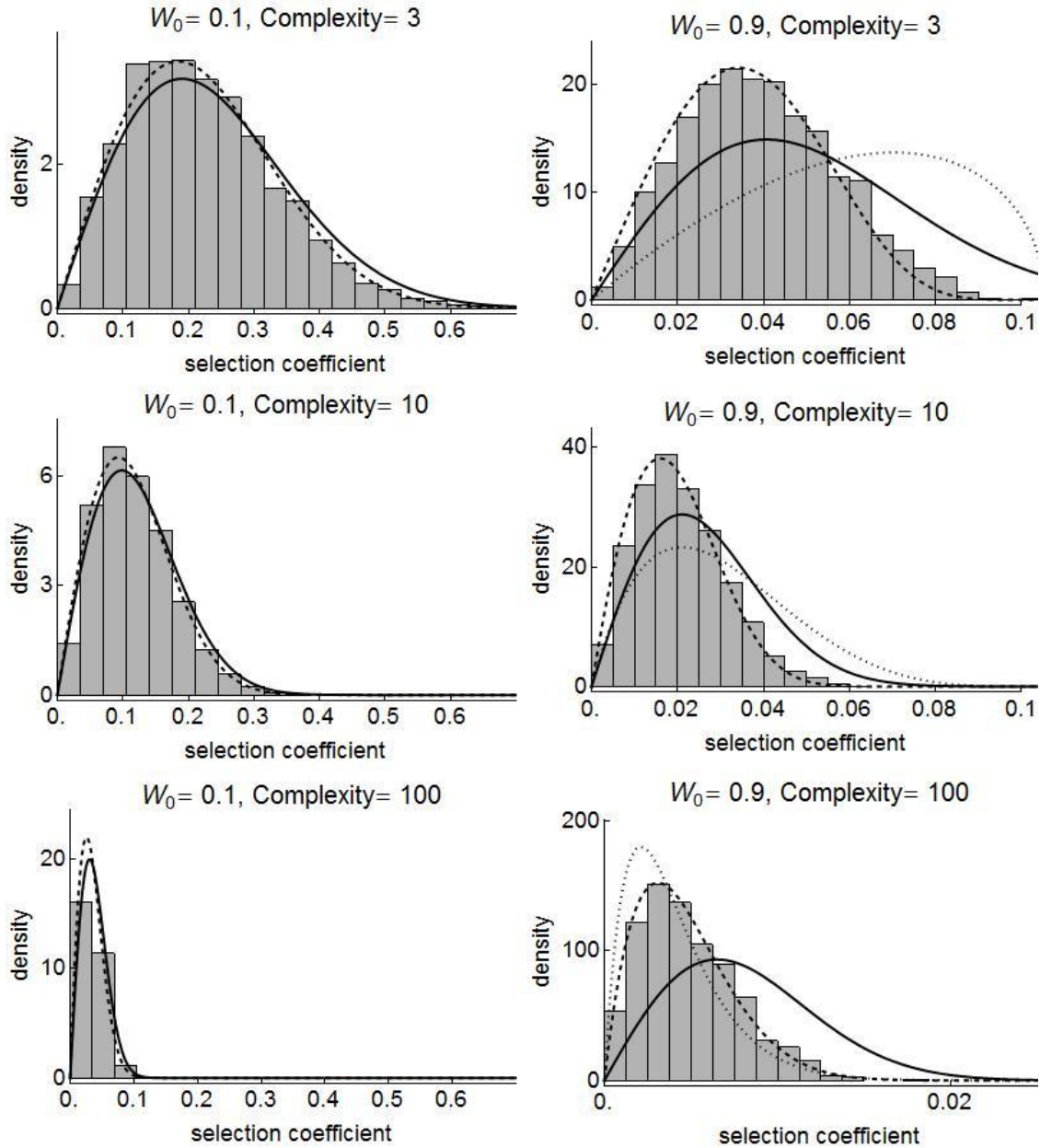

Figure 2. Distribution of the selection coefficient *s* under Fisher's model for various fitnesses of the ancestral strain $W_0$ and complexities of the phenotypic space. The plain line is the $\chi_2$ analytical approximation (equation 2), the dashed line is based on the gamma approximation developed in Martin and Lenormand (2006) (see Appendix) and the dotted line for $W_0 = 0.9$ is the beta approximation based on extreme value theory, developed in Martin and Lenormand (2008). Selection coefficient is calculated for **10000** selected mutations. $\sigma_{mut}$ is scaled such that the average norm of the mutational effect on phenotype is constant equal to 0.1 across complexities. The population size is $N = 10^7$ and the mutation rate $\mu = 10^{-9}$.

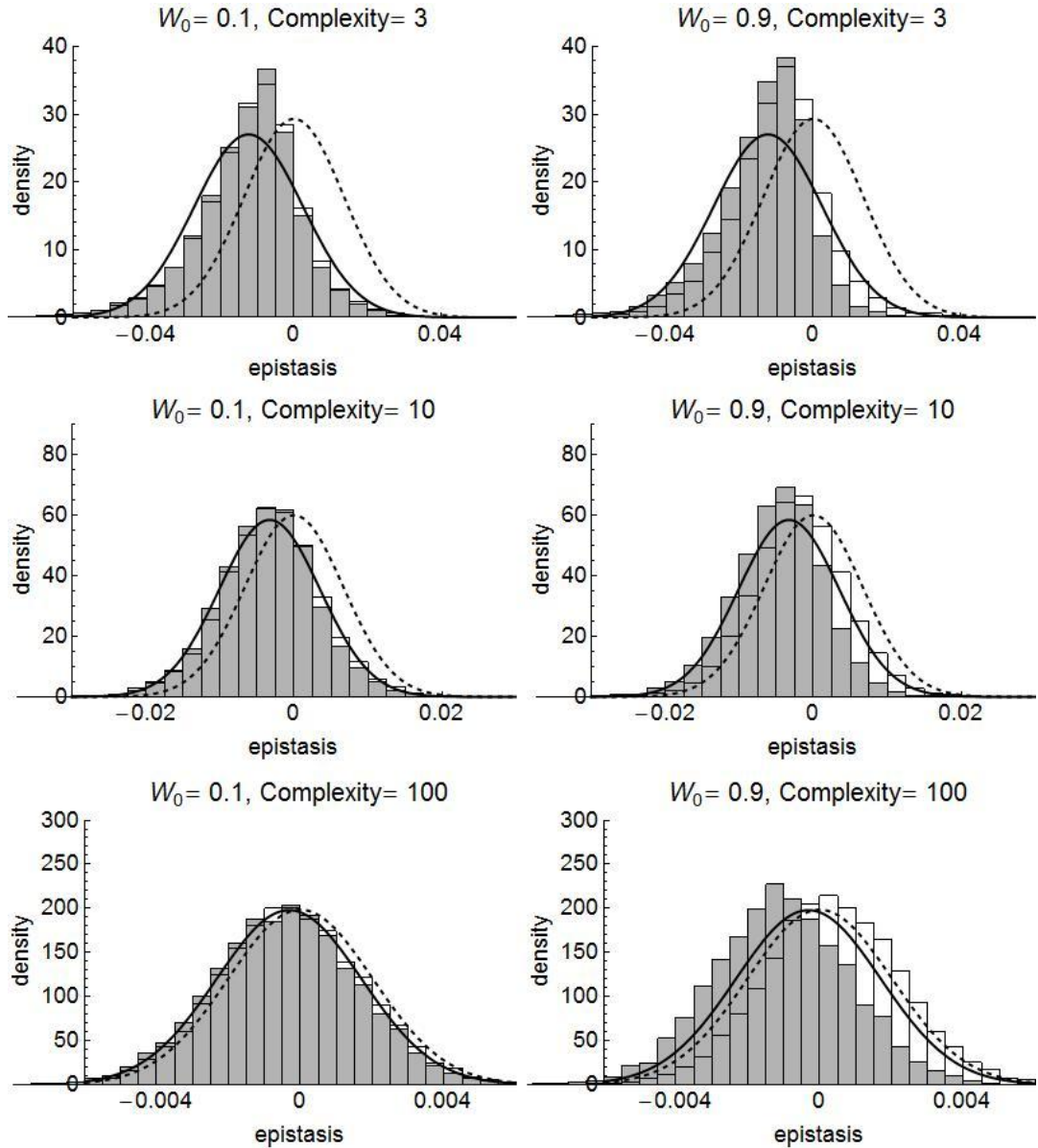

Figure 3. Distribution of the epistasis coefficient *e* between two independently selected mutations (grey) and for co-selected mutations (white) for various fitnesses of the ancestral strain $W_0$ and complexities of the phenotypic space. The plain line is the analytical approximation for independently selected mutations and the dashed line is the normal approximation for random (newly arising) mutations (Martin *et al.* 2007). For independently selected mutations, the first mutations sweeping through the population in each of $20000$ independent replicates were selected, and epistasis coefficient is calculated for $10000$ independent pairs of selected mutations. For co-selected mutations, the first two mutations sweeping through the population in each of $10000$ independent replicates were selected, resulting in $10000$ independent epistasis coefficients. $\sigma_{mut}$ is scaled such that the average norm of the mutational effect on phenotype is constant equal to 0.1 across complexities. The population size is $N = 10^7$ and the mutation rate $\mu = 10^{-9}$.

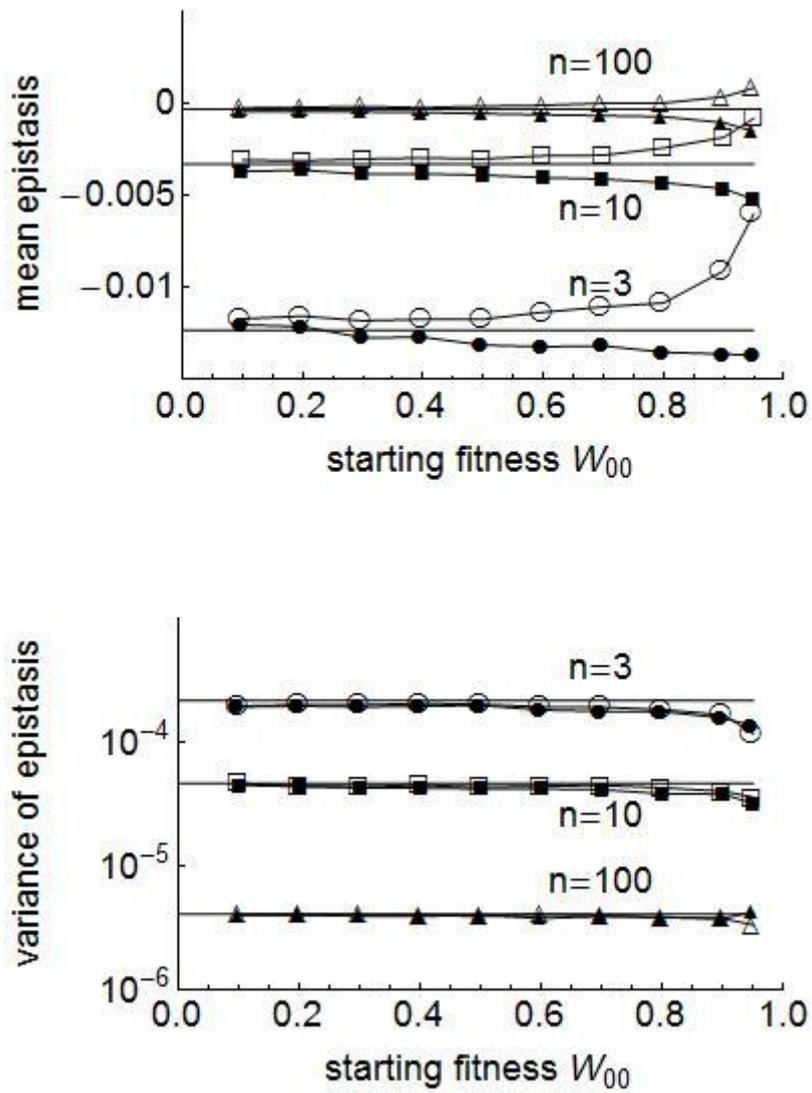

Figure 4. Mean (top graph) and variance (bottom graph) of the epistasis coefficient $e$ as a function of the fitness of the ancestral strain, between two independently selected mutations (closed symbols) and between two co-selected mutations (open symbols). This is shown for complexities of the phenotypic space $n = 3$ (circles), 10 (squares) and 100 (triangles). The plain line is the analytical approximation for independently selected mutations. The epistasis coefficient is calculated among at least 2000 independent pairs of mutations. Other parameters as in fig. 3.

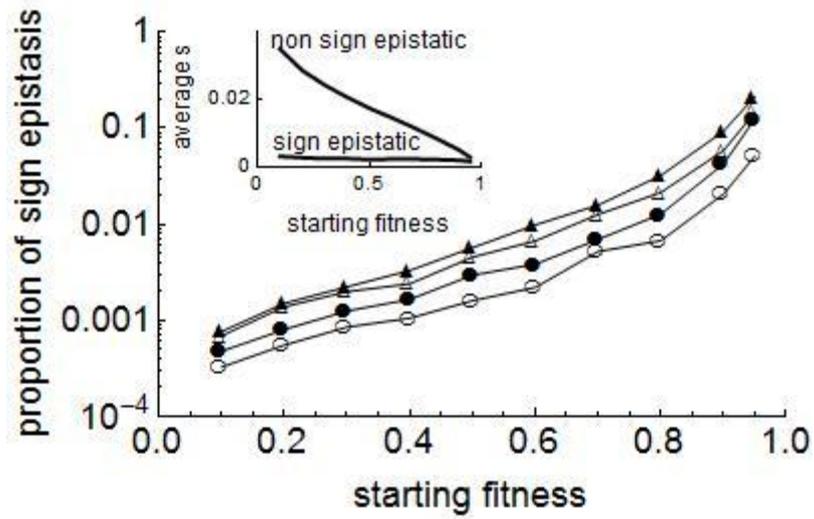

Figure 5: Fraction of sign epistasis between two independently selected mutations (filled symbols) and between two co-selected mutations (open symbols). This is shown for complexities of the phenotypic space $n = 3$ (circles) and 100 (triangles). Inset shows the average coefficient of selection among non sign epistatic mutations (top curve) and sign epistatic mutations (bottom curve), as a function of the starting fitness, for complexity $n = 3$ and independently selected mutations (curves are similar for other parameters and selection procedure). The fraction of sign epistasis is calculated among at least 2000 independent pairs of mutations. Other parameters as in fig. 3.

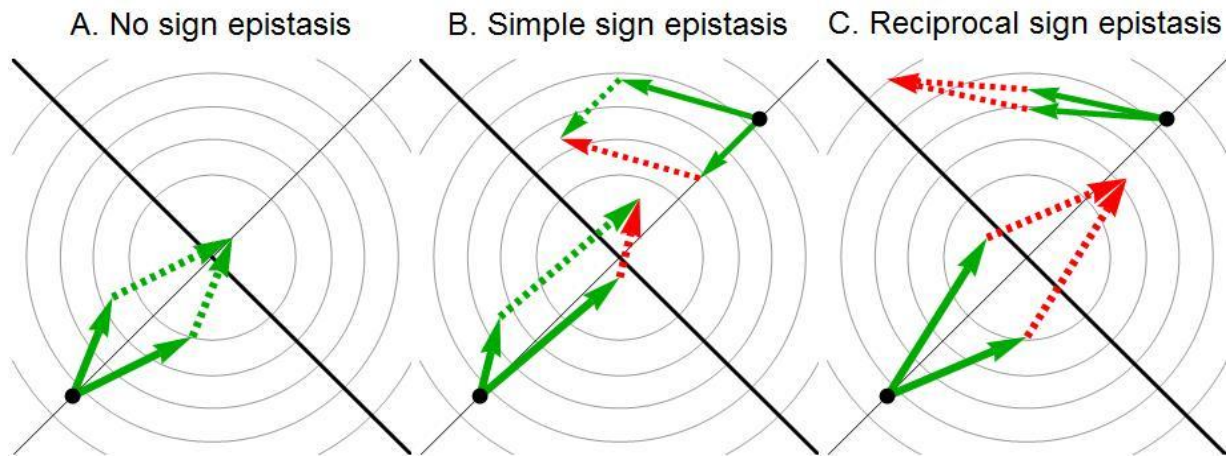

Figure 6: The geometry of sign epistasis among two mutations in a two-dimensional Fisher's fitness landscape model. The light gray lines are the fitness isoclines and the black lines are the phenotypic axes. Beneficial and deleterious mutations are shown respectively as green and red arrows in the phenotypic space. In panel A, the two mutations are beneficial in the ancestral background and in the background with the other mutation (no sign epistasis). In panels B and C, two examples of pairs of sign epistatic mutations are shown. In B, one of the mutations is deleterious in the background with the other mutation (simple sign epistasis). In C, both mutations are deleterious in the background with the other mutation (reciprocal sign epistasis).

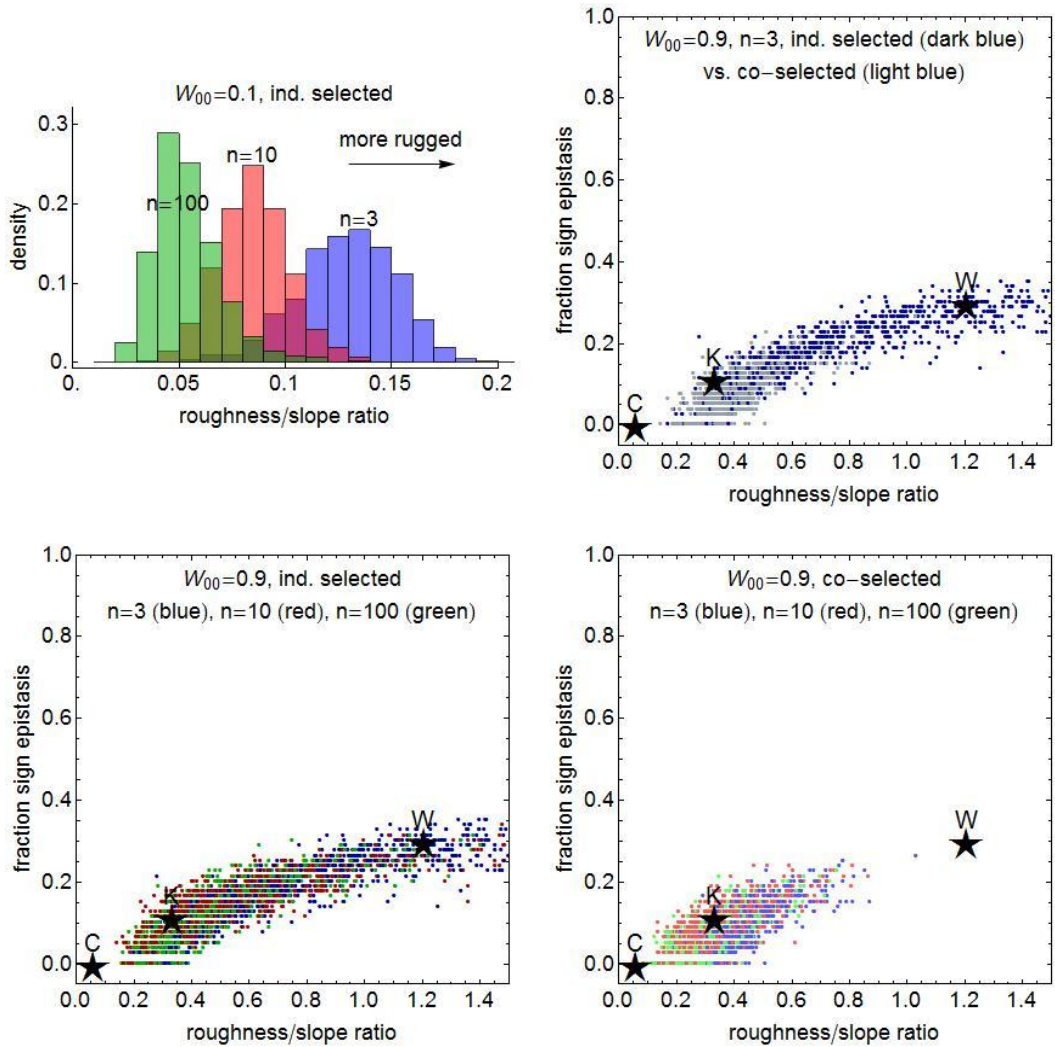

Figure 7: The distribution of the roughness to slope ratio and the fraction of sign epistasis over 1000 genotypic fitness landscapes generated with 5 mutations. Top left panel: distribution of roughness to slope ratio when $W_{00} = 0.1$, for three levels of complexity, for independently selected mutations (distributions for co-selected mutations are very similar). In these conditions, there is no sign epistasis in more than 95% of the landscapes. Top right panel: distribution of the statistics in independently selected (dark blue) vs. co-selected mutations (light blue). Bottom panel: distribution of the statistics for different levels of complexity ($n = 3, n = 10, n = 100$ in blue, red, green) in independently selected mutations (left) and co-selected mutations (right). Statistics corresponding to three experimental landscapes are superimposed (C: Chou et al. 2011, K: Khan et al. 2011, W: Weinreich et al. 2006).

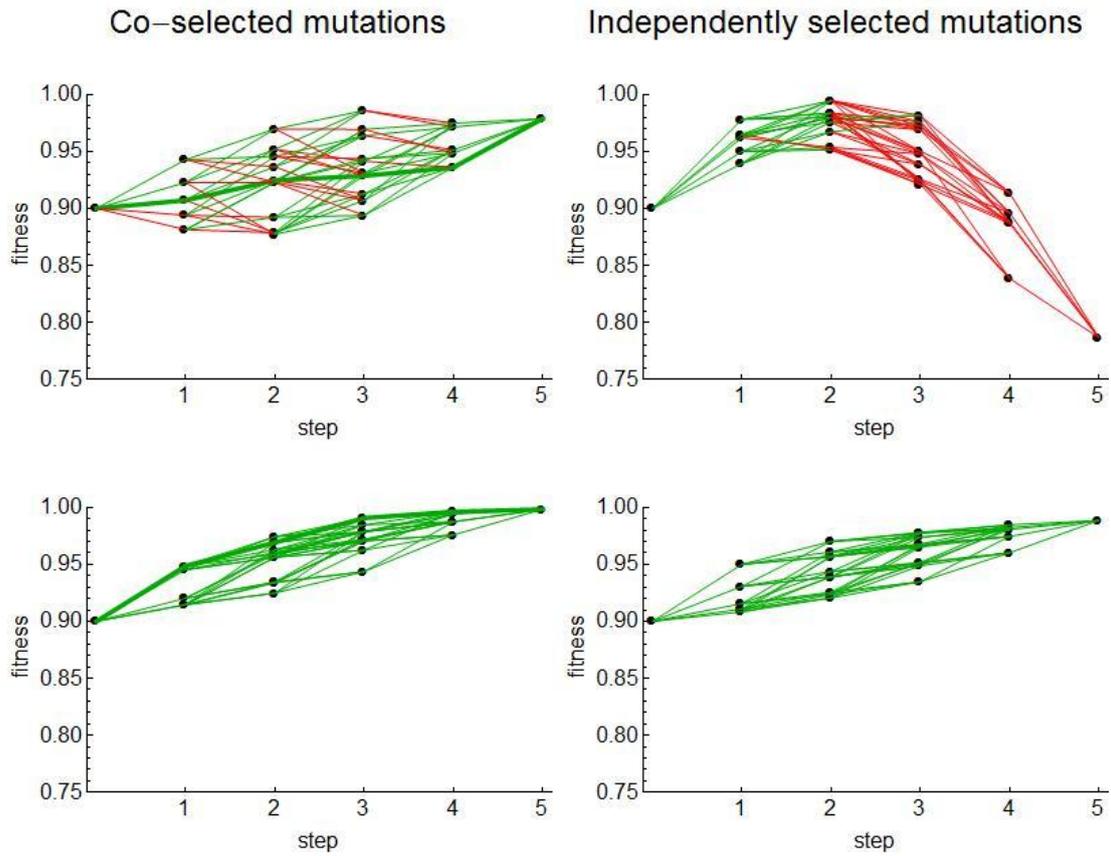

Figure 8: Rough (top) and smooth (bottom) fitness landscapes obtained with 5 co-selected (left) or independently selected (right) mutations in Fisher's model, with the same parameters ($n = 3, W_{00} = 0.9$). Fitness of the $2^5 = 32$ genotypes is shown as a function of the step in the adaptive walk. The black points represent genotypes' fitnesses, and the green and red links are beneficial and deleterious mutations respectively. The thicker links represent the evolutionary path that was actually taken in the simulation, for co-selected mutations.

# Appendix for "Properties of selected mutations and emergence of complex genotypic landscapes under Fisher's Geometric Model"

## 1. Properties of selected mutations

Distribution of phenotypic effects and selection coefficient of fixed mutations

We use the following fitness function to define the relationship between phenotypes and fitness: $W[z] = \exp[-\sum_{i=1}^{n} z_i^2]$, where $z_i$ is the phenotype $i$ and $n$ is the total number of phenotypes (the complexity). Without loss of generality, we assume the phenotype of the ancestral strain is $\{\sqrt{-\log[W_{00}]}, 0, 0 \ldots, 0\}$ where $W_{00}$ is the fitness of the ancestral strain. We make the approximation that selection operates only along the first phenotypic axis (this approximation works best when the ancestral strain is far from the optimum). As a consequence, selection affects the distribution of phenotypic effects of *selected* mutations along the first axis only, and the distribution of phenotypic effects along all other $n - 1$ directions is exactly the same as the distribution of *random* phenotypic effects (i.e., $N(0, \sigma_{mut}^2)$ where $N$ stands for the normal distribution).

To derive the distribution of selected phenotypic effects along the first axis, we assume we are in a "strong selection – weak mutation" regime. In this regime, the next beneficial mutation to fix in the population is obtained by a random drawing among the pool of beneficial mutations, with each mutation is weighted by its selection coefficient (Kimura 1983). Thus, the distribution of *selected* phenotypic effect will be the distribution of *random* phenotypic effect weighted by the selective effect of the mutation along the first axis. To derive the distribution of phenotypic effects along the first axis, we need to know what is the selection coefficient acting on a mutant with effect on the first trait $dz_1$. This is given by:

$$s_1 = \underbrace{-(z_{0,1} + dz_1)^2}_{\substack{log-fitness \\ of\, mutant}} + \underbrace{z_{0,1}^2}_{\substack{log-fitness \\ of\, wildtype}} = -2z_{0,1}dz_1 - dz_1^2 \qquad (S1)$$

with $z_{0,1} = \sqrt{-\log[W_{00}]}$. Thus, the distribution of phenotypic effects along the first direction is given by the product of the selective coefficient $-2z_{0,1}dz_1 - dz_1^2$ and the distribution of phenotypic effects of random mutations $e^{\frac{-dz_1^2}{2\sigma_{mut}^2}}$ (Fisher 1930, Kimura 1983), that is:

$$g(dz_1) = \frac{1}{\lambda}\left(-2z_{0,1}dz_1 - dz_1^2\right)e^{\frac{-dz_1^2}{2\sigma_{mut}^2}} \text{ if } dz_1 < 0$$

$$g(dz_1) = 0 \text{ otherwise} \qquad (S2a)$$

where $\lambda$ is a normalizing constant. Similarly the distribution of fitness effect of fixed mutations is given by:

$$f^*(s) = \frac{1}{\kappa}sf(s) \text{ if } s > 0$$

$$f^*(s) = 0 \text{ otherwise} \qquad (S2b)$$

where $\kappa$ is a normalizing constant and $f(s)$ is the distribution of fitness effects of *random* mutations.

Assuming mutation effects on the phenotype are small relative to the distance to the optimum, we may ignore the $dz_1^2$ term in (S2a), and the selection coefficient is directly proportional to the phenotype at the first axis $s \approx -2z_{0,1}dz_1$. In this case the distribution of phenotypic effects of mutations along the first axis simplifies to:

$$g(dz_1) = \frac{-dz_1}{\sigma_{mut}^2} e^{\frac{-dz_1^2}{2\sigma_{mut}^2}} \text{ if } dz_1 < 0$$

$$g(dz_1) = 0 \text{ otherwise}$$

(S3a)

The above expression reveals that $-d\, z_1/\sigma_{mut}$ follows a $\chi_2$ ($\chi_i$ denotes a chi distribution with $i$ degrees of freedom). Similarly the distribution of selection coefficients of selected mutations is:

$$f^*(s) = \frac{s}{4\sigma_{mut}^2(-\log[W_{00}])} e^{\frac{-s^2}{8\sigma_{mut}^2(-\log[W_{00}])}} \text{ if } s > 0$$

$$f^*(s) = 0 \text{ otherwise}$$

(S3b)

The selection coefficient normalized by $4\sigma_{mut}^2(-\log[W_{00}])$ follows a $\chi_2$. Note that several others approximations for $f(s)$ may be plugged into equation (S2b) to find other approximations for the fitness effects of selected mutations (e.g., the normal distribution of Waxman and Peck 1998, Lourenço et al. 2011, or the gamma distribution of Martin and Lenormand 2006). For example, we used the displaced gamma distribution proposed by Martin and Lenormand and obtained the following distribution:

$$f^*(s) = \frac{1}{\kappa} s \frac{1}{\Gamma(\beta)\alpha^\beta} (-s - \log[W_{00}])^{\beta-1} e^{\frac{--s-\log[W_{00}]}{\alpha}} \text{ if } 0 < s < -\log[W_{00}]$$

$$f^*(s) = 0 \text{ otherwise}$$

(S3c)

where $\alpha = \frac{2(1+2\epsilon)\sigma_{mut}^2}{1+\epsilon}$, $\beta = \frac{n(1+\epsilon)^2}{2+4\epsilon}$ with $\epsilon = \frac{-\log[W_{00}]}{n\sigma_{mut}^2}$ and $\kappa$ is a normalizing constant (with a lengthy expression). We also compared these analytical predictions with the beta distribution based on Extreme Value Theory (Martin and Lenormand 2008), given by:

$$f^*(s) = s \frac{\left(1 + \frac{s}{\log[W_0]}\right)^{\frac{n}{2}-1}}{\beta\left[2, \frac{n}{2}\right] \log[W_0]^2} \text{ if } 0 < s < -\log[W_0] \quad \text{(S3d)}$$

$$f^*(s) = 0 \text{ otherwise}$$

Geometry of selected mutations

The algebraic results derived above can be translated in geometric terms. The angle between a selected mutation and the first phenotypic axis can be calculated as:

$$\theta = \tan^{-1}\left[\frac{\sqrt{\sum_{i=2}^{n} dz_i^2}}{\sqrt{dz_1^2}}\right] \quad \text{(S4)}$$

where the numerator is the norm of the resultant vector of the phenotypic space in all "neutral directions (phenotypic directions 2 to $n$), and the denominator is the norm of the vector in the first phenotypic direction (under selection). Because both the numerator and the denominator follow $\chi$ distributions when appropriately scaled, the quantity $\frac{2}{n-1}\tan[\theta]^2$ is distributed according to a F-distribution with degrees of freedom $n-1$ (corresponding to the numerator) and $2$ (corresponding to the denominator) when appropriately scaled. It follows a relatively simple expression for the distribution of the angle $\theta$:

$$f(\theta) = (n-1)\cos[\theta]\sin^{n-2}[\theta] \text{ for } \theta \in \left[0, \frac{\pi}{2}\right] \quad \text{(S5)}$$

Lastly, the norm of *selected* mutations scaled by $\sigma_{mut}$ is distributed according to a chi distribution with degrees of freedom $(n+1)$. This follows from the fact that the effect along the first trait normalized by $\sigma_{mut}$ is $\chi_2$ while all other effects (along the $n-1$ other traits) are $N(0, \sigma_{mut}^2)$ distributed. Thus the sum of square is a $\chi_{n+1}^2$ and the norm is a $\chi_{n+1}$.

These geometrical results can be extended to describe the relationship between two independently selected mutations. Whatever the complexity of the phenotypic space, it is possible to represent the relationship between two independent mutations and the optimum in a 3-dimensional space. This can be done using the Gram-Schmidt process, which generates an orthonormal basis for this 3-dimensional space in which the norm and angles are conserved (note that this space will of course be different for each pair of mutations considered). The relative disposition of the two mutations in the 3D space is characterized by their norms, the angles they have with the main direction of selection, and the angle between the two mutations when projected on a plane orthogonal to the main axis of selection (the azimuth). Because the fitness effect of a mutation does not change when it revolves around the main axis of selection, this angle is well characterized by the distribution of angles of random mutations for $n-1$ phenotypic directions, that is (Poon and Otto 2000):

$$f(\psi) = \frac{\Gamma\left(\frac{n-1}{2}\right)}{\sqrt{\pi}\Gamma\left(\frac{n-2}{2}\right)} \sin[\psi]^{n-3} \text{ for } \psi \in [0, \pi] \quad \text{(S6)}$$

The angles of the two mutations with the main axis of selection, which we call $\theta$ and $\theta'$ together with the azimuth $\psi$ are sufficient to find the angle between two mutations $\alpha$:

$$\cos[\alpha] = \sin[\theta]\sin[\theta']\cos[\psi] + \cos[\theta]\cos[\theta'] \tag{S7}$$

where $\theta$ and $\theta'$ lie in $\left[0, \frac{\pi}{2}\right]$ and $\psi$ in $[0, \pi]$. Although we know the distribution of $\theta$, $\theta'$ and $\psi$, we were not able to derive explicitly the distribution of $\alpha$. The average of the distribution of $\cos[\alpha]$ is $\frac{\pi}{4} \frac{\Gamma\left(\frac{n+1}{2}\right)^2}{\Gamma\left(1+\frac{n}{2}\right)^2}$. This average is a decreasing function which is equal to $4/9$ when $n = 3$ and tends to 0 as the complexity increases. Hence, the distribution of $\cos[\alpha]$ becomes increasingly concentrated around 0 as the complexity increases (i.e., the distribution of $\alpha$ is concentrated around $\pi/2$). Thus, in a very complex phenotypic space two independent mutations tend to be perpendicular in the phenotypic space.

## 2. Distribution of epistasis among selected mutations

Epistasis among two mutations is defined as:

$$e = \log\left[\frac{W_{11} W_{00}}{W_{01} W_{10}}\right] \tag{S8}$$

Where $W_{00} = W_0$ is the fitness of the ancestral strain, $W_{11}$ is the fitness of the double mutant, and $W_{10}$ and $W_{01}$ are the fitnesses of the two single mutants. Under Fisher's model of adaptation where the fitness of a genotype characterized by $n$ traits $z_i$ with $i \in [1, n]$ is given by $e^{-\sum_{i=1}^{n} z_i^2}$, epistasis reduces to (Martin et al. 2007):

$$e = -2 \sum_{i=1}^{n} dz_i dz_i' \tag{S9}$$

where $dz_i$ and $dz_i'$ are the phenotypic effects of two independent mutations that evolved in this background.

Under our approximation, we can partition epistasis into a component due to mutational effect along the first (selected) axis and a component due to mutational effect along all other axes:

$$e = \underbrace{-2dz_1dz_1'}_{e_{sel}} \underbrace{-2\sum_{i=2}^{n} dz_i dz_i'}_{e_{neutral}} \tag{S10}$$

The second component follows the distribution of neutral epistasis for $n-1$ traits. This can be approximated by a normal distribution with mean $0$ and variance $4(n-1)\sigma^4$ (Martin *et al.* 2007). The first component is the product of two independent random variables following the distribution given by (S2a). It can be shown using the standard formula for the probability density function of two independent random variables that the probability density function of this component is:

$$p(e_{sel}) = \frac{-e_{sel}}{4\sigma_{mut}^4} Y_0\left(\frac{-e_{sel}}{2\sigma_{mut}^2}\right) \text{ if } e_{sel} < 0$$

$$p(e_{sel}) = 0 \text{ if } e_{sel} > 0 \tag{S11}$$

where $Y_0(.)$ is a Bessel function of the second kind (it is defined as the solution of a differential equation). The total density of epistasis is thus a convolution between the function defined in (S11) and the density of a $N(0, 4(n-1)\sigma_{mut}^4)$:

$$h(e) = \int_{-\infty}^{0} \left(\frac{1}{2\sqrt{2\pi}\sqrt{n-1}\sigma_{mut}^2} \exp\left[\frac{-(e-e_{sel})^2}{2*4(n-1)\sigma_{mut}^4}\right]\right)\left(\frac{-e_{sel}}{4\sigma_{mut}^4} Y_0\left[\frac{-e_{sel}}{2\sigma_{mut}^2}\right]\right) de_{sel} \tag{S12}$$

We were not able to find a simpler expression for this convolution. But as the number of traits increases, epistasis should increasingly look like a normal distribution with proper mean and variance, because all non-selected traits will progressively have more weight in the convolution. The mean and the variance of epistasis can be found by summing up the mean and variances of the two components of epistasis. The mean and variance of selected epistasis are:

$$E[e_{sel}] = -\pi\sigma_{mut}^2$$

$$V[e_{sel}] = (16 - \pi^2)\sigma_{mut}^4 \tag{S13}$$

Epistasis corresponding to the selected component is always negative. The mean and variance of total epistasis are:

$$E[e] = -\pi\sigma_{mut}^2$$

$$V[e] = (4n + 12 - \pi^2)\sigma_{mut}^4 \tag{S14}$$

**3. Sign epistasis among independently selected mutations**

For independently selected mutations, the relationships $W_{10} > W_{00}$ and $W_{01} > W_{00}$ must hold in a strong selection, weak mutation regime, because the mutations are both beneficial in the ancestral strain. Thus, the first mutation is sign epistatic if and only if $W_{11} < W_{01}$ (an analogous condition holds for the second mutation). This condition is equivalent to:

$$\boxed{\begin{array}{l}\log(W_{11}) - \log(W_{01}) < 0 \text{, equivalent to :}\\[2mm]
\underbrace{\log(W_{11}) + \log(W_{00}) - \log(W_{01}) - \log(W_{10})}_{e} + \underbrace{\log(W_{10}) - \log(W_{00})}_{s_{(1)}} = e + s_{(1)}\\[2mm]
\qquad < 0\end{array}} \quad (S15)$$

Thus, a selected mutation is sign epistatic with another mutation, relative to the ancestral background, if and only if the sum of its selective coefficient and its epistasic coefficient is negative.

For co-selected mutations, the relationships $W_{10} > W_{00}$ and $W_{11} > W_{10}$ must hold. Thus the first mutation is sign epistatic if and only if $W_{11} < W_{01}$, and the second mutation is sign epistatic if and only if $W_{01} < W_{00}$. The first condition is equivalent to the condition expressed in (S15) and the second condition is equivalent to:

$$\boxed{\begin{array}{l}\log(W_{01}) - \log(W_{00}) < 0 \text{, equivalent to :}\\[2mm]
\underbrace{-\log(W_{11}) - \log(W_{00}) + \log(W_{01}) + \log(W_{10})}_{-e} + \underbrace{\log(W_{11}) - \log(W_{10})}_{s_{(2)}}\\[2mm]
\qquad = -e + s_{(2)} < 0\end{array}} \quad (S16)$$

where $s_{(2)}$ denotes the selection coefficient of the second mutation. Note that reciprocal sign epistasis cannot happen between co-selected mutations, because it would imply that $W_{11} < W_{00}$.

## References of Appendix